\documentclass[aps,showpacs,nofootinbib]{revtex4}
\usepackage{bm}
\usepackage{graphics}
\usepackage{epsfig}
\usepackage{graphicx}
\usepackage{amsmath}
\usepackage{amsfonts}
\usepackage{amssymb}
\begin{document}

\title{Phenomenological theory of nonergodic phenomena in dipole- and spin-glasses}
\author{P. N. Timonin}
\email{timonin@aaanet.ru}
\affiliation{Physics Research Institute at
Southern Federal University, 344090 Rostov-on-Don, Russia}

\date{\today}

\begin{abstract}
The path-dependent magnetizations (polarizations) and susceptibilities in dipole- and spin-glasses are described analytically for the standard protocols of temperature and field variations using the phenomenological Landau-type description of multiple metastable states in the nonergodic phases. The immediate manifestation of metastable states' multiplicity as staircases of magnetizations (polarizations) curves in the temperature cycling experiments is explained and described. The obtained results are in a reasonable qualitative agreement with the existing experimental data.    
\end{abstract}


\maketitle

\section{Introduction}
\label{intro}

The notorious property of spin- and dipole-glass phases in disordered magnets and ferroelectrics is the dependence of a sample state on its previous history, that is on its path in the field-temperature plane to a specific point. Thus arriving to some point at the plane where such phase exists via different protocols (paths), say, field-cooled (FC) and zero-field-cooled (ZFC) ones, we get two different states of a glass system with different net moments, susceptibilities etc. And seemingly both these states are quite stable - on the laboratory time scale (hours and days) they show no tendency to relax to some unique stable state, see, for example, Refs. \cite{1}, \cite{2}. In general using the variety of the paths to a specific field-temperature point we can get in it a number of such presumably stable states with quite diverse thermodynamic parameters and having different degree of order in the local moment's directions.

Such behavior strongly implies that the phase space of the glassy system is divided into separate regions by the very high free energy barriers so the Arrhenius times for the transitions between these regions are much greater than the laboratory times. Now it is not known if this means that the true nonergodicity sets in the spin- and dipole-glass phases, i. e. if the free energy barriers diverge in the thermodynamic limit so the system can be confined in the separate "ergodic component" of phase space for arbitrarily long time in sufficiently large samples. In such a case  the quasi-static evolution of a glassy system can be described within the notion of "metastable states" - local free energy minima each belonging to the unique ergodic component.  

Such nonergodic picture of the glassy phases with multiple metastable states naturally arises in the mean-field spin-glass models \cite{3}. In this framework the seemingly nonergodic behavior of random magnets and ferroelectrics can be qualitatively reproduced in numerical simulations of the mean-field models \cite{4}-\cite{7}. It should be noted that this agreement with the present day experiments having time spans up to $10^5$ sec can fail at longer times as it is quite possible that there are actually only finite barriers in real glassy phases and realistic short-range models. Then the restoration of the ergodicity at longer times would manifest itself by the slow convergence of the numerous presumably metastable states to a true stable one after, say, a month or a year even if the influence of external random fields is eliminated in the experiment. Yet till now we have neither definite theoretical nor experimental evidences in favor of finite or divergent barriers.
  
In spite of this uncertainty in the nature of irreversible phenomena in the glassy phases of disordered magnets and ferroelectrics the mean-field paradigm of their nonergodicity can be very useful for (at least qualitative) description of these phenomena on the time scales of the order $10^5 -10^6$ sec and, probably, even longer ones. The attractiveness of this approach lies in its simplicity and visual clarity. Here the quasi-static evolution of nonergodic system is associated with the free energy landscape in which it moves between a number of local minima as this landscape changes under field and temperature variations. 

The phenomenological form of such free energy describing the competition between the ferro-phase and the glassy one was suggested in Ref. \cite{8}. This phenomenology is based on the observation \cite{9} that in the mean-field approach the mechanism of phase transition in random system can be viewed as a result of the condensation of some collective excitations (modes) represented by the delocalized eigenvectors of the matrix of random exchange (in magnets) or random matrix of force constants (in random ferroelectrics). In strongly disordered short-range systems the modes which can condense first at the glass transition are necessarily those lying at the localization threshold \cite{9}, \cite{10}. According to the random matrix theory \cite{10} in various short-range random matrix ensembles the modes near this threshold  have specific sparse fractal structure. It appears that the number of sites participating in them is not proportional to the full number of sites $N$ as usual but only to $N^x, x < 1$. This means that the condensation of one such mode results in the appearance of local spontaneous moments in a small fraction of the sample's sites - on the sparse fractal set spreading throughout all sample (as the mode is delocalized) and having the fractal dimension $d_f = xd < d$, $d$ is the sample's dimension. It is argued in Ref. \cite{9} that the condensation of one such mode can not stabilize the other fractal modes which do not overlap with the already condensed one. So the condensation of such modes will proceed until almost all sites of a sample acquire spontaneous moments. Apparently it needs macroscopic number $N_0=N/N^x=N^{1-d_{f}/d}$ of modes to cover all sites. As the condensation temperatures of the fractal modes are determined by the corresponding eigenvalues which are close to the localization threshold and differ by the order $1/N$ values \cite{10} the temperature interval in which the sequence of their condensations would take place can be quite small - of the order $N_0*1/N = N^{-d_{f}/d}$.

Thus the phenomenological potential of Ref. \cite{8} is the function of $N_0$ magnetizations (polarizations) $m_i$ of these fractal modes which describe the prevailing orientations of spins (dipole moments) in different nonoverlapping fractal sets of sites consisting of approximately $N_1 = N^x = N^{d_{f}/d}$ sites. It can have up to $2^{N_0}$ local minima (metastable states) which differ by the values and signs of  $m_i$. It appears that the transitions between these minima involves necessarily the change of sign of at least one $m_i$ and this needs to overcome the potential barrier of the order $N_1$. Thus we have the divergent in the thermodynamic limit barriers but it can be just the consequence of the mean-field character of this phenomenology allowing only the simultaneous upturn of $N_1$ spins. 

Meanwhile the less energy consuming paths between the local minima may exist in realistic short-range models. Nowadays we have the numerical results for $3d$ short-range Edwards-Anderson Ising spin-glass certifying that the collective excitations in it do have fractal character with $d_f \approx 2.10$ and with the energy cost diminishing with their size \cite{11}, \cite{12}. The latter point out to the absence of divergent barriers yet one should be cautious extrapolating the results of simulations for $N=10^3$ to the real systems with $N \sim 10^{18}$. Also one may tentatively suppose that sequential upturns of the fractal set of $N_1$ spins on the way between two local minima would create the domain wall of a sort between the upturned spins and non-upturned ones with the fractal dimension $d_f -1$ when $d_f >1$. The energy cost of such wall is of the order $N^{(d_f-1)/d}$ so we still have the divergent barrier between minima albeit much lower than in the mean-field potential. Further numerical simulations on larger samples may shed some light on the validity of such arguments for short-range systems.
 
With these reservations we may turn to the phenomenological mean-field description of the widely used standard protocols (such as FC and ZFC ones) for quasi-static irreversible processes in the spin- and dipole-glass phases. We use the basic qualitative feature of the metastable states phenomenology of Ref. \cite{8} for these phases - the existence of inclined hysteresis loop filled with magnetization (polarization) curves so the sides of loop are the stability limits of metastable states. We show that the difference of the thermodynamic parameters in standard protocols can be explained by the trapping of a system in different metastable states resulting from the hysteresis loop's temperature evolution. The obtained results allow to explain qualitatively the wealth of experimental data on the temperature and field dependencies of net moment in standard protocols \cite{13} - \cite{21} and in their temperature-cycling modifications resulting in the "staircases" of magnetization (polarization) curves \cite{2}, \cite{16}, \cite{22}.

\section{Metastable states and hysteresis loops}
\label{sec:1}
First we briefly recall the essential results of Ref. \cite{8} using the magnetic terminology. According to Ref. \cite{8} the phenomenological potential $F$ for the randomly frustrated uniaxial ferromagnet can be expressed via the magnetizations $m_i$, $i=1$,…,$N_0$, of sparse fractal modes 
\[
F\left( {\bf{m}} \right) = \frac{{\tau _g }}{2}\left[ {m^2 } \right] + \frac{{\tau _f  - \tau _g }}{2}\left[ m \right]^2  + \frac{a}{4}\left[ {m^4 } \right] + \frac{b}{4}\left[ {m^2 } \right]^2  - h\left[ m \right]
\]
Here $\left[ {m^k } \right] = N_0^{ - 1} \sum\limits_{i = 1}^{N_0} {m_i^k }$,  $N_0  =  N^{1 - \left( {d_f /d} \right)}$. The coefficients $a>0$ and $b>0$ in $F$ are some constants specific for a given disorder realization, while $\tau_f$ and $\tau_g$ are linear decreasing functions of temperature $T$ changing their signs at temperatures $T_f $ and $T_g$ correspondingly also being disorder dependent.

This $F\left( {\bf m} \right)$ is symmetric under all permutations of $m_i$ and describes the competition of ferromagnetic order with the spin glass one represented by the  $N_0  - 1$ - dimensional order parameter composed of the independent $m_i - m_j$ components favoring the antiparallel $m_i$ orientations. This potential have up to $2^{N_0}$ metastable states with partially ordered $m_i$ in either fully ordered (ferromagnetic) ($m_i = m_0$) or spin-glass phase ($\left[ {m} \right] = 0$). In zero field the ordinary thermodynamics predicts the transition between these phases at $\tau_f =\tau_g$. We consider here only the spin-glass region $\tau_g<0$, $\tau_g <\tau_f $.
In a specific crystal $\tau_f$ is some function of $\tau_g$ which is a linear one at small $\tau_g$. So we introduce
\[
t \equiv \tau _g /a,\qquad t'  \equiv \tau _f /a = t_0  + ct,\qquad t_0  > 0,
\qquad c < 1.
\]

The conditions on the parameters $t_0$ and $c$ imply the presence of only one transition into the glass phase at $t=0$ in zero field $h=0$ after which the crystal always stays in this phase at $t<0$. So the homogeneous fully ordered state with $m_i=m_0$ for all $i$ is only the metastable one at $t<0$. Meanwhile in the paraphase at $t>0$ it is the only stable state. Its magnetization $m_0$ obeys the equation 
\[	
h/a = t' m_0  + \left( {1 + \beta } \right)m_0^3 ,\qquad \beta  \equiv b/a.
\]
and it is stable for
\[
t + \left( {3 + \beta } \right)m_0^2  > 0
\]

In the inhomogeneous metastable states appearing at $t<0$ $m_i$ can acquire just two values: $m_+>0$ and 
\begin{equation}
m_ -   =  - xm_ +   < 0,
\label{eq:1}
\end{equation}

So they can be characterized by the number $N_+$ of $m_+$ values in it or by the parameter
\[
n = N_ +  /N_0, \qquad 0 < n < 1
\]

All thermodynamic parameters of a state with a given $n$ can be obtained as functions of $n$ and the parameter $x$ defined in Eq. (\ref{eq:1}). Thus \cite{8}
\begin{eqnarray*}
m_ +   = \sqrt { - t} R\left( {x,n} \right),
\\
R\left( {x,n} \right) = \left\{ {1 - x + x^2  + \beta \left[ {n + \left( {1 - n} \right)x^2 } \right]} \right\}^{ - 1/2} 
\end{eqnarray*}

and the net magnetization of such state is
\begin{eqnarray}
m = M\left( {x,n,t} \right),
\label{eq:2}
\\
M\left( {x,n,t} \right) = \sqrt { - t} \left[ {n - \left( {1 - n} \right)x} \right]R\left( {x,n} \right).
\nonumber
\end{eqnarray}
The expression for the susceptibility 
\[
\chi  = \frac{{\partial m\left( {n,t,h} \right)}}{{\partial h}},
\]
of a state with a given $n$ is

\begin{equation}
\chi ^{ - 1} \left( {x,n,t} \right)a^{ - 1}  = t'  - t\left[ {1 + R\left( {x,n} \right)^2 \frac{{\left( {1 + x} \right)\left( {2x - 1} \right)\left( {2 - x} \right) + 2\beta \left[ {\left( {2x - 1} \right)n + x^2 \left( {2 - x} \right)\left( {1 - n} \right)} \right]}}{{\left( {2x - 1} \right)n + \left( {2 - x} \right)\left( {1 - n} \right) + 2\beta n\left( {1 - n} \right)\left( {1 + x} \right)}}} \right].
\label{eq:3}
\end{equation}

\begin{figure*}
\centering
\includegraphics{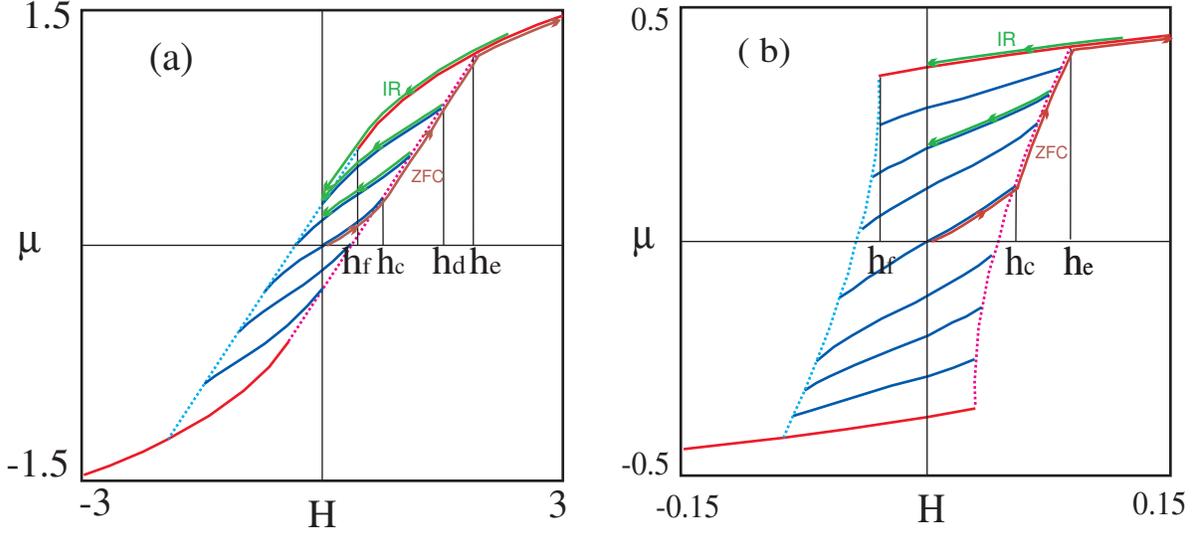}
\caption{(color online) The field dependenciese of magnetizations of metastable states with different $n$; (a) $t=-0.5$, $t'=0.25$, $\beta =0.1$, (b) $t=-1.5$, $t'=-1.25$, $\beta =5$. Directed lines show the system's evolution in ZFC and IR processes.}\label{Fig.1}
\end{figure*}

The parameter $x = x(n,t,h)$ can be found from the equation of state of the form
\begin{eqnarray}
H\left( {x,n,t} \right) = h,
\label{eq:4}
\\
H\left( {x,n,t} \right)/a = \left( {t'  - t} \right)M\left( {x,n,t} \right) + \sqrt { - t^3 } x\left( {1 - x} \right)R\left( {x,n} \right)^3 
\nonumber
\end{eqnarray}
The stability condition for a metastable state in glassy phase reads
\begin{equation}
1/2 < x\left( {n,t,h} \right) < 2
\label{eq:5}
\end{equation}
Thus the knowledge of $x = x(n,t,h)$ provides us with the full description of the thermodynamics of metastable states and the regions of their existence. Yet to graph the field dependencies of their magnetizations we do not need to solve Eq. (\ref{eq:4}) as the couple of Eqs. (\ref{eq:2}, \ref{eq:4}) gives the parametric representations of them with the parameter $x$ varying between the stability boundaries of Eq. (\ref{eq:5}). In the glass phase considered these $m=m(n, t, h)$ are shown in Fig. \ref{Fig.1} for some values of $t, t'$ and $\beta$. They do not cross one another and fill the interior of inclined hysteresis loop. Such loops are often seen in disordered magnets and ferroelectrics, see, for example, Refs. \cite{23}-\cite{25}.

As we see the ends of $m=m(n, t, h)$ curves at which metastable states loose their stability may be thought of as the upper and lower branches of the loop while the top and the bottom of the loop are bounded by the $m_0(h)$ curve.
The stability lines $x=2$ and $x=0.5$ representing these upper and lower branches on ($m, h$) plane are described by the following parametric equations
\[
\begin{array}{l}
 m_u \left( {h,t} \right):m_u  = M\left( {2,n,t} \right),{\rm{  }}h = H\left( {2,n,t} \right),{\rm{  }}0 < n < 1 \\ 
 m_l \left( {h,t} \right):m_l  = M\left( {0.5,n,t} \right),{\rm{  }}h = H\left( {0.5,n,t} \right),{\rm{  }}0 < n < 1 \\ 
 \end{array}
\]
Excluding the parameter $n$ from them we find
\begin{eqnarray*}
h/a = \left( {t'  - t} \right)m_u  - \frac{1}{4}\left( {\frac{{\beta m_u  + \sqrt {\beta ^2 m_u^2  - 4t\left( {3 + 2\beta } \right)} }}{{3 + 2\beta }}} \right)^3, 
\\
 - h_e  < h < h_f \left( { - m_e  < m_u  < m_f } \right){\rm{,  }} x = 2;
\\
h/a = \left( {t'  - t} \right)m_l  - \frac{1}{4}\left( {\frac{{\beta m_l  - \sqrt {\beta ^2 m_l^2  - 4t\left( {3 + 2\beta } \right)} }}{{3 + 2\beta }}} \right)^3, 
\\
 - h_f  < h < h_e \left( { - m_f  < m_l  < m_e } \right){\rm{,  }} x = 0.5.
\end{eqnarray*}
Here the parameters
\begin{eqnarray*}
h_f  = H\left( {2,1,t} \right) = a\sqrt {\frac{{ - t}}{{3 + \beta }}} \left( {t'  - t\frac{{1 + \beta }}{{3 + \beta }}} \right),
\\
m_f  = M\left( {2,1,t} \right) = \sqrt {\frac{{ - t}}{{3 + \beta }}}; 
\\
h_e  = H\left( {0.5,1,t} \right) = 2a\sqrt {\frac{{ - t}}{{3 + 4\beta }}} \left( {t'  - 4t\frac{{1 + \beta }}{{3 + 4\beta }}} \right),
\\
m_e  = M\left( {0.5,1,t} \right) = 2\sqrt {\frac{{ - t}}{{3 + 4\beta }}}. 
\end{eqnarray*}
define the points at which $m_u$ and $m_l$ join the $m_0(h)$ curve, see Fig. \ref{Fig.1}.
The useful characteristics of the loop are also the field and magnetization of the most disordered state with $n=0.5$ at the point in which $m(0.5,t,h)$ joins the lower branch, see Fig. \ref{Fig.1}:
\[
\begin{array}{l}
h_c  = H\left( {0.5,0.5,t} \right) = a\sqrt {\frac{{ - t}}{{12 + 10b}}} \left( {t'  - t\frac{{14 + 5\beta }}{{6 + 5\beta }}} \right),
\\
m_c  = M\left( {0.5,0.5,t} \right) = \sqrt {\frac{{ - t}}{{12 + 10\beta }}}. 
\end{array}
\]

\begin{figure*}
\centering
\includegraphics{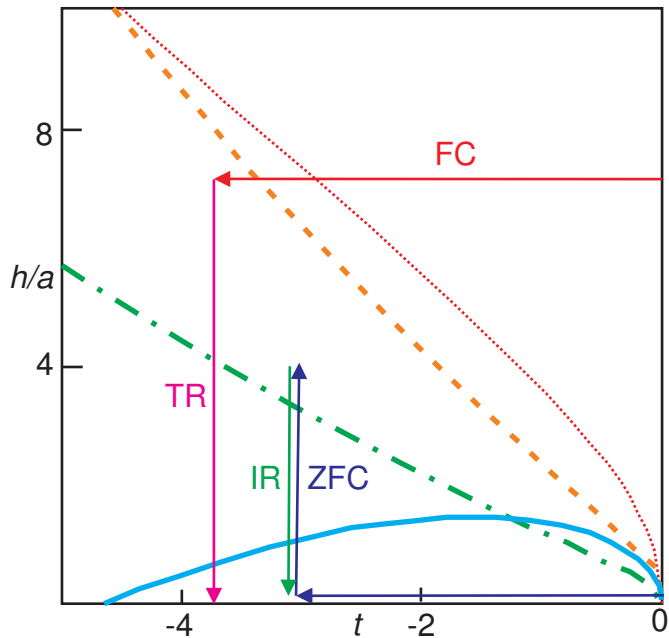}
\caption{(color online) The temperature dependence of characteristic fields:  $h_f$ (solid line), $h_e$ (dotted line), $h_d$ (dashed line), $h_c$ (dash-dotted line) for $\beta = 0.1$, $c = 1$, $t_0 = 3$. Directed lines show the temperature and field variations for standard protocols.}\label{Fig.2}
\end{figure*}

For the temperature region where $h_f >0$ we can also define the parameter $n_d(t)$ of the state having the left stability boundary at zero field, see Fig. \ref{Fig.1}(a). From the equation $H\left( {2,n_d ,t} \right) = 0$
we have
\begin{equation}
n_d \left( t \right) = 1 + \frac{1}{{2\beta }} - \sqrt {\left( {\frac{1}{{2\beta }} + \frac{1}{3}} \right)^2  + \frac{{2t}}{{9\beta \left( {t'  - t} \right)}}} 
\label{eq:6}
\end{equation}
The right boundary field of this state is
\begin{equation}
h_d /a \equiv H\left( {0.5,n_d ,t} \right) = \frac{{9\left( {1 + \beta } \right)\left( {2n_d  - 1} \right)\left( {t'  - t} \right)\sqrt { - t} }}{{\left[ {3 + \beta \left( {3n_d  + 1} \right)} \right]^{3/2} }}
\label{eq:7}
\end{equation}
The temperature behavior of these fields is shown in Fig. \ref{Fig.2}. Note that for the $c<1$ case considered here they all stay positive except of $h_f$ which can become negative if
\begin{equation} 
c > c_0  = \frac{{1 + \beta }}{{3 + \beta }}
\label{eq:8}
\end{equation}                                                                   at 
\begin{equation} 
t < t_{f0}  \equiv  - \frac{{t_0 }}{{c - c_0 }}.                                 \label{eq:9}
\end{equation}

Some basic properties of the glassy phase can be deduced from Fig. \ref{Fig.2}. Thus above $h_e$ the system has unique stable state - the homogeneous one. Below $h_f$ this state ceases to exist and only some disordered states stay stable here. 
  
\section{Standard protocols: ZFC and IR}
\label{sec:2}
The values of the introduced characteristic fields play important role in the determination of the results of the standard protocols used in experimental studies of dipole- and spin-glasses. The paths in ($h, t$) plane for these protocols are shown in Fig. \ref{Fig.2}. Along with the field-cooled (FC) and zero-field-cooled (ZFC) paths there are also thermo-remanent (TR) and isothermal-remanent (IR) ones which can be seen as the extensions of FC and ZFC via the last step - the switching off the field. Mostly the TR and IR protocols are used to obtain the magnetizations - the so-called TRM and IRM. 

Here we should note that many experimental studies do not strictly follow the standard ZFC, IR and TR protocols, often after turning on (off) the field the crystal is heated and the parameters recorded during the heating are thought to be the ZFC, IR and TR ones \cite{1}. Generally it may not be so, thus obtained data may differ from that of the standard procedures and may depend on the temperature at which the cooling ends up. Fortunately, in the glass phase such heating procedures give the data identical to the standard ZFC and TR protocols as we shall see later. But it is not true for IR protocol.

Every path (protocol) in Fig. \ref{Fig.2} is uniquely defined by some $t, h$ values and now we can find the dependence of their resulting thermodynamic parameters on these $t, h$. Noticing that in real experiments and simulations the magnetization follows the outline of the inclined hysteresis loop under quasi-static variations of large-amplitude field we may conclude that under the quasi-static conditions:

a)	the system does not leave the metastable state until it becomes unstable;

b)	leaving the unstable state the system goes to the metastable state with the nearest net magnetization.

These rules seem to be the almost apparent consequences of Langevin dynamics for our potential $F\left( {\bf{m}} \right)$ and they allow to easily reproduce the quantitative features of the above protocols in a glassy phase. It is quite easy to find $m_{ZFC}$ as cooling at zero field brings crystal to the non-magnetized and most deep \cite{8} state with $n=0.5$. Then depending on the magnitude of applied (positive) field the system may stay at this state or enter the lower branch of the hysteresis loop or acquire the $m_0$ value, see Fig. \ref{Fig.1}. So we have
\begin{equation} 
m_{ZFC}  = \left\{ \begin{array}{l}
 m_g ,{\rm{  }}h < h_c , \\ 
 m_l ,{\rm{  }}h_c  < h < h_e , \\ 
 m_0 ,{\rm{  }}h_e  < h. \\ 
 \end{array} \right.
\label{eq:10}
\end{equation}
Here $m_g$ is the magnetization of $n=0.5$ state defined in parametric form as
\begin{eqnarray}
m_g \left( {h,t} \right):m_g  = M\left( {x_g ,0.5,t} \right),{\rm{  }}h = H\left( {x_g ,0.5,t} \right), \label{eq:11}
\\
0.5 < x_g  < 1.
\nonumber
\end{eqnarray}
Excluding the parameter $x_g$ we find the equation for $m_g$ in the form
\begin{eqnarray}
h/a = \left( {t'  - t\frac{{3 + \beta }}{{1 + \beta }}} \right)m_g  - 4\frac{{2 + \beta }}{{1 + \beta }}m_g^3 ,
\label{eq:12}
\\
{\rm{    }}\left| h \right| < h_c {\rm{  }}\left( {\left| {m_g } \right| < m_c } \right).
\nonumber
\end{eqnarray}
Also using the above considerations on the evolution of $x$ and $n$ parameters in ZFC process we can find $\chi_{ZFC}$ from Eqs. (\ref{eq:3}, \ref{eq:4}). Thus we have
\begin{eqnarray}
\chi _{ZFC}  = \left\{ \begin{array}{l}
 \chi \left( {x_g ,0.5,t} \right), \qquad h < h_c , \\ 
 \chi \left( {0.5,n_r ,t} \right),\qquad h_c  < h < h_e , \\ 
 \chi _0 ,\qquad h_e  < h. \\ 
 \end{array} \right.
\label{eq:13}
\\
\begin{array}{l}
H\left( {0.5,n_r ,t} \right) = h,
\\
\chi _0  = a^{ - 1} \left[ {t'  + 3\left( {1 + \beta } \right)m_0^2 } \right]^{ - 1}
\end{array}
\label{eq:14} 
\end{eqnarray}

Here $n_r=n_r(h,t)$ is the parameter of the state to which system arrives at the lower loop's branch.

The IR process proceeds in the loop interior as shown in Fig. \ref{Fig.1} and its description is not more difficult. Evidently $m_{IR}$ (IRM) stays zero at $h<h_c$ as then the system does not leave the $n=0.5$ state.  At higher fields $m_{IR}$ is the zero-field magnetization of the state with $n=n_r$ which system join at ZFC process, see Eq. (\ref{eq:14}). The result of the further field growth depends on the $h_f$ sign at a given $t$. If $h_f>0$ and $h>h_d$ then system ends up on the upper branch of the loop after switching off the field (Fig. \ref{Fig.1}a) while for $h_f<0$ and $h>h_e$ it occurs at the fully ordered state (Fig. \ref{Fig.1}b). Thus
\begin{eqnarray}
m_{IR}  = \left\{ \begin{array}{l}
 m_r \vartheta \left( {h_d  - h} \right)\vartheta \left( {h - h_c } \right)
\\
 + m_u \left( {h = 0} \right)\vartheta \left( {h - h_d } \right),{\rm{  }}h_f  > 0 
\\ 
 m_r \vartheta \left( {h_e  - h} \right)\vartheta \left( {h - h_c } \right) 
 \\
 + m_0 \left( {h = 0} \right)\vartheta \left( {h - h_e } \right){\rm{,  }}h_f  < 0 \\ 
 \end{array} \right.
\label{eq:15}
\end{eqnarray}
\[
\begin{array}{l}
m_r  = M\left( {x_r ,n_r ,t} \right)
\\
x_r \left( {h,t} \right): H\left( {x_r ,n_r ,t} \right) = 0.
\\
m_0 \left( {h = 0} \right) = \sqrt {\frac{{ - t' }}{{1 + \beta }}} ,  
\end{array}
\]
\begin{equation}
m_u \left( {h = 0} \right) = M(2,n_d,t) = \left( {3n_d  - 2} \right)^{3/2} \sqrt {\frac{{t'  - t}}{2}} 
\label{eq:16}
\end{equation}
Here $\vartheta (x)$ is the Heaviside's step function.

Similarly we get the susceptibility for this protocol
\begin{eqnarray}
\chi _{IR}  = \left\{ \begin{array}{l}
 \chi \left( {1,0.5,t} \right)\vartheta \left( {h_c  - h} \right) 
 \\
 + \chi \left( {x_r ,n_r ,t} \right)\vartheta \left( {h_d  - h} \right)\vartheta \left( {h - h_c } \right) 
 \\
 + \chi _u \left( {h = 0} \right)\vartheta \left( {h - h_d } \right),{\rm{  }}h_f  > 0 
 \\ 
 \chi \left( {1,0.5,t} \right)\vartheta \left( {h_c  - h} \right) 
 \\
 + \chi \left( {x_r ,n_r ,t} \right)\vartheta \left( {h_e  - h} \right)\vartheta \left( {h - h_c } \right)
 \\
  + \chi _0 \left( {h = 0} \right)\vartheta \left( {h - h_e } \right){\rm{,  }}h_f  < 0 \\ 
 \end{array} \right.
\label{eq:17}
\\ 
\begin{array}{l}
 \chi _u \left( {h = 0} \right) = \chi \left( {2,n_d ,t} \right) = \frac{{1 + 2\beta \left( {1 - n_d } \right)}}{{a\left( {1 + \beta n_d } \right)\left( {t'  - t} \right)}}  
\\
 \chi \left( {1,0.5,t} \right) = a^{ - 1} \left( {t'  - t\frac{{3 + \beta }}{{1 + \beta }}} \right)^{ - 1} ,
\\
 \chi _0 \left( {h = 0} \right) =  - \left( {2at' } \right)^{ - 1}.
\end{array}  
\label{eq:18}
\end{eqnarray}

\section{Standard protocols: FC and TR}
\label{sec:3}

\begin{figure*}
\centering
\includegraphics{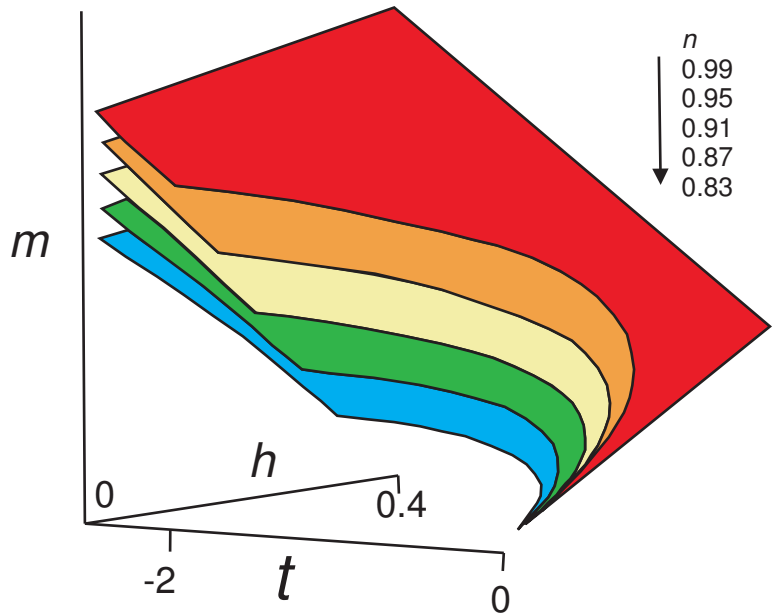}
\caption{(color online) The magnetizations of metastable states $m(n, t, h)$ for $\beta = 0.5$, $c=0.9$, $t_0=1$.}\label{Fig.3}
\end{figure*}

To describe the FC and TR parameters we should consider the relative positions of the boundaries of metastable magnetizations $m=m(n, t, h)$ defined by Eqs. (\ref{eq:2}, \ref{eq:4}). They are shown for several $n$ in Fig. \ref{Fig.3} and we can see that the edges of $m(n, t, h)$ sheets form something like an amphitheater (with the infinitesimal steps of order $N_0^{-1}$).
\begin{figure*}
\centering
\includegraphics{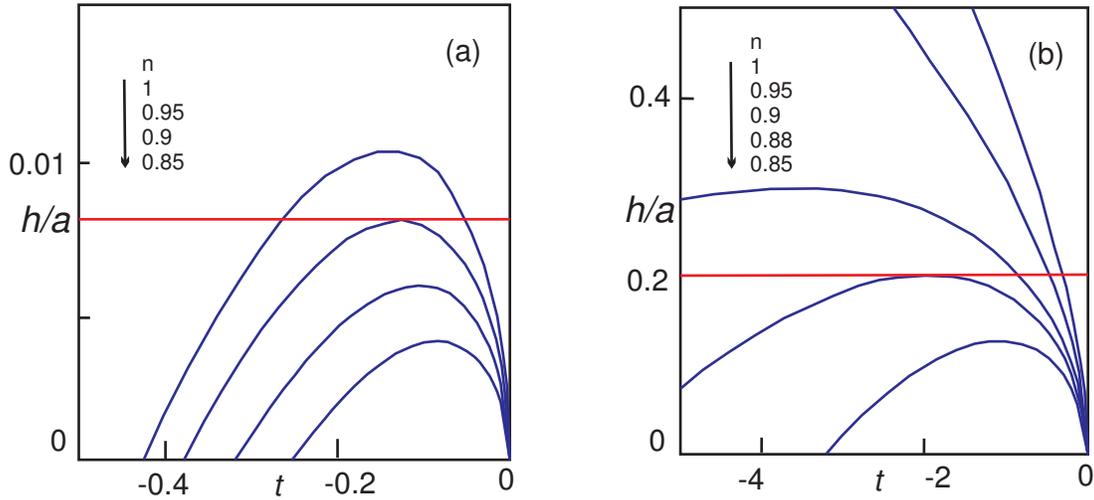}
\caption{(color online) The $x=2$ stability lines of metastable states for (a) $\beta=3$, $c=0.9$, $t_0=0.1$, ($c>c_0$) and (b) $\beta=1$, $c=0.4$, $t_0=0.7$, ($c<c_0$). The states with given $n$ are stable above these lines.}\label{Fig.4}
\end{figure*}
So when cooling in constant field (in FC protocol) the system comes to the boundary of the fully ordered state the sequence of infinitesimal jumps to the states with lower $n<1$ begins. This proceeds until the system joins the state with some $n_s$ which is stable for all $t$ at given $h$. Then on further cooling the system will be trapped in this state. From Fig. \ref{Fig.4} showing projections of the $x=2$ stability lines of $m(n, t, h)$ sheets on the ($h, t$) plane we can conclude that this last state has the stability line maximum touching the line $h=const$. So to find the parameter $n=n_s$ for this state and the value $ t=t_s$ at which the system becomes trapped in it we should solve the following equations
\begin{equation}
n_s \left( h \right),{\rm{ }}t_s \left( h \right):{\rm{  }}H\left( {2,n_s ,t_s } \right) = h,{\rm{ }}\frac{{\partial H\left( {2,n_s ,t_s } \right)}}{{\partial t_s }} = 0.
\label{eq:19}
\end{equation}

The solutions to Eqs. (\ref{eq:19}) can be represented as
\begin{equation}
n_s  = \frac{{2 + z}}{3},\qquad t_s  =  - \frac{{t_0 }}{3} \cdot \frac{{z\left( {3 + 2\beta  - \beta z} \right)}}{{2 - \left( {1 - c} \right)z\left( {3 + 2\beta  - \beta z} \right)}}
\label{eq:20}
\end{equation}
where $z$ obeys the equation
\begin{eqnarray}
z^3  + \left( {\frac{h}{{2a}}} \right)^2 \left( {\frac{3}{{t_0 }}} \right)^3 \left[ {\left( {1 - c} \right)z\left( {3 + 2\beta  - \beta z} \right) - 2} \right] = 0,\label{eq:21}
\\
0 < z < 1. {\rm{   }}
\nonumber
\end{eqnarray}
Once we have this $z$ we can find the magnetization for the $n_s$-state which system joins on cooling at $t < t_s$
\[
m_s \left( {h,t} \right):m_s  = M\left( {x,n_s ,t} \right),{\rm{  }}h = H\left( {x,n_s ,t} \right),{\rm{  1}} < x < 2.
\]
Then FC magnetization is
\begin{equation}
m_{FC}  = \left\{ \begin{array}{l}
 m_0 ,{\rm{  }}t_s  < t,{\rm{  }}h_f  < h, \\ 
 m_u ,{\rm{  }}t_s  < t < 0,{\rm{  }}h < h_f , \\ 
 m_s ,{\rm{  }}t < t_s . \\ 
 \end{array} \right.
\label{eq:22}
\end{equation}
This result is valid when FC path cross the stability lines of metastable states . It is always so if $c <c_0$ but for $c > c_0$ it is only true for fields lower than the maximum reached by $h_f$, cf. Fig. \ref{Fig.4}a.
\[
h_{f\max }  = 2a\sqrt {\frac{{t_0 ^3 }}{{27\left( {3 + \beta } \right)\left( {c - c_0 } \right)}}}. 
\]

So for $c > c_0 $ Eq. (\ref{eq:22}) is valid at $0 < h < h_{f\max } $ only.
If $c > c_0 ,{\rm{ }}h > h_{f\max } $ the system always stays in the fully ordered state so 
\begin{equation}
m_{FC}  = m_0
\label{eq:23}
\end{equation}                                                          
In the same way we get for $c < c_0$ or $c > c_0$, $h < h_{f\max } $
\begin{equation}
\chi _{FC}  = \left\{ \begin{array}{l}
 \chi _0 ,{\rm{  }}t_s  < t,{\rm{  }}h_f  < h, \\ 
 \chi _u ,{\rm{  }}t_s  < t < 0,{\rm{  }}h < h_f , \\ 
 \chi _s ,{\rm{  }}t < t_s . \\ 
 \end{array} \right.
\label{eq:24}
\end{equation} 
with
\begin{eqnarray*} 
\chi _u  = \chi \left( {2,n_u ,t} \right),{\rm{   }}H\left( {2,n_u ,t} \right) = h,{\rm{  }}2/3 < n_u  < 1
\\
\chi _s  = \chi \left( {x,n_s ,t} \right),{\rm{  }}H\left( {x,n_s ,t} \right) = h,{\rm{  }}1 < x < 2
\end{eqnarray*} 
Evidently, for $c > c_0$, $h > h_{f\max } $
\begin{equation}
\chi _{FC}  = \chi _0 
\label{eq:25}
\end{equation} 
To get the $m_{TR}$ (TRM) we should consider the effect of the switching off the field on the FC state. Turning to Fig. \ref{Fig.1} we conclude that if FC state becomes unstable at $h=0$ then system ends up on the upper branch of the loop, otherwise it preserves the state (the $n$ number) down to $h=0$. As
\[
H\left( {2,n_s ,3t_s } \right) = 0
\]
the last happens for $n_s$ state at $t<3t_s$ while the fully ordered state can stay stable down to $h=0$ at $t<t_{f0}$ when $c>c_0$, see Eqs. (\ref{eq:8}, \ref{eq:9} ). Thus we have
for $c < c_0$ or $c > c_0$, $h < h_{f\max }$ 
\begin{eqnarray}
m_{TR}  = \left\{ \begin{array}{l}
 m_u \left( {h = 0} \right),{\rm{  3}}t_s  < t < 0, \\ 
 m_{s0} ,{\rm{  }}t < 3t_s . \\ 
 \end{array} \right.
\label{eq:26}
 \\
\begin{array}{l}
 m_{s0} \left( {h,t} \right) = M\left( {x_0 ,n_s ,t} \right),{\rm{    }} \\ 
 x_0 \left( {h,t} \right){\rm{: }}H\left( {x_0 ,n_s \left( h \right),t} \right) = 0. \\ 
 \end{array}
\nonumber
\end{eqnarray}
and for $c > c_0$, $h > h_{f\max }$ 
\begin{equation}
m_{TR}  = \left\{ \begin{array}{l}
 m_u \left( {h = 0} \right),{\rm{  }}t_{f0}  < t < 0,{\rm{  }}\left( {h_f  > 0} \right), \\ 
 m_0 \left( {h = 0} \right),{\rm{  }}t < t_{f0} {\rm{,  }}\left( {h_f  < 0} \right). \\ 
 \end{array} \right.
 \label{eq:27}
\end{equation}

For $\chi_{TR}$ we obtain for $c < c_0$ or $c > c_0$, $h < h_{f\max }$ 
\begin{equation}
\chi _{TR}  = \left\{ \begin{array}{l}
 \chi _u \left( {h = 0} \right),{\rm{  3}}t_s  < t < 0, \\ 
 \chi \left( {x_0 ,n_s ,t} \right),{\rm{  }}t < 3t_s . \\ 
 \end{array} \right.
 \label{eq:28}
\end{equation}
and for $c > c_0$, $h > h_{f\max }$ 
\begin{equation}
\chi _{TR}  = \left\{ \begin{array}{l}
 \chi _u \left( {h = 0} \right),{\rm{  }}t_{f0}  < t < 0,{\rm{  }}\left( {h_f  > 0} \right), \\ 
 \chi _0 \left( {h = 0} \right) =  - \left( {2at' } \right)^{ - 1} ,{\rm{  }}t < t_{f0} {\rm{,  }}\left( {h_f  < 0} \right). \\ 
 \end{array} \right.
 \label{eq:29}
\end{equation}

\begin{figure*}
\centering
\includegraphics{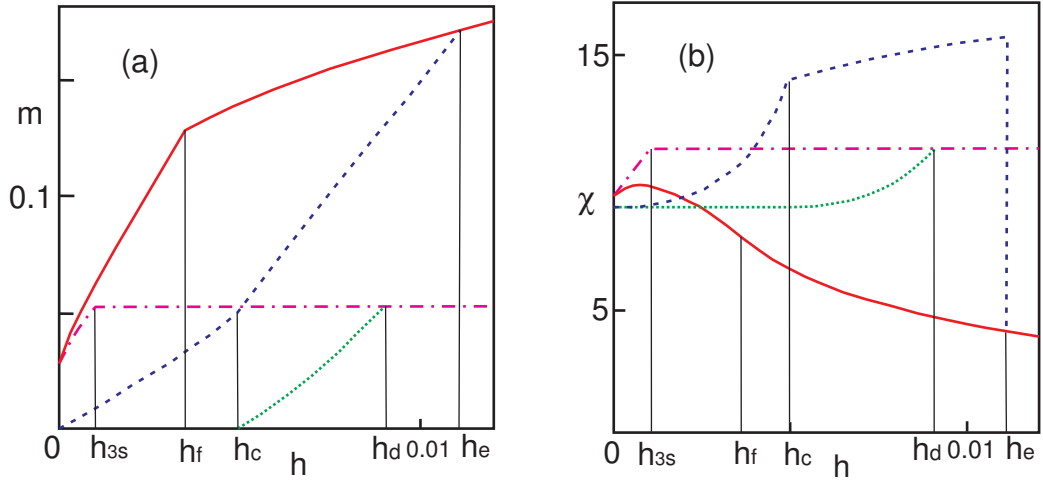}
\caption{(color online) Field dependencies of magnetizations (a) and susceptibilities (b) for FC (solid lines), ZFC (dashed lines), IR (dotted lines) and TR (dash-dotted lines) protocols for $a=1$, $\beta=2$, $c = 0.9$, $t_0 = 0.05$, $t=-0.08$.}\label{Fig.5}
\end{figure*}

\begin{figure*}
\centering
\includegraphics{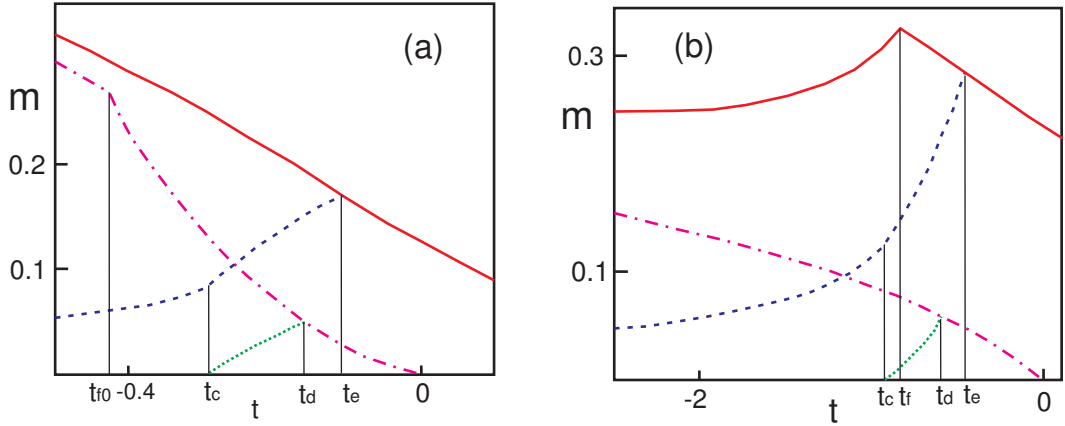}
\caption{(color online) Temperature  dependencies of magnetizations for FC (solid lines), ZFC (dashed lines), IR (dotted lines) and TR (dash-dotted lines) protocols for (a)$\beta =3$, $c = 0.9$, $t_0 = 0.1$, $h=0.02$,  $c>\beta/(2+\beta)$  and (b)$\beta =5$, $c = 0.5$, $t_0 = 0.1$, $h=0.1$,  $c<\beta/(2+\beta)$.}\label{Fig.6}
\end{figure*}

\begin{figure*}
\centering
\includegraphics{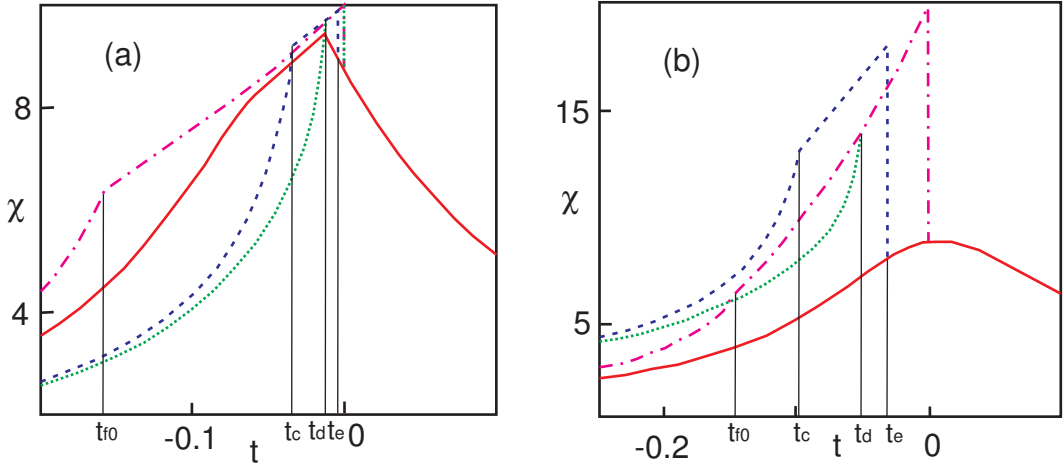}
\caption{(color online) Temperature  dependencies of susceptibilities for FC (solid lines), ZFC (dashed lines), IR (dotted lines) and TR (dash-dotted lines) protocols for (a)$\beta=0.5$, $c = 0.9$, $t_0 = 0.1$, $h=0.006$ and (b)$\beta =2$, $c = 0.9$, $t_0 = 0.05$, $h=0.006$. $a=1$.}\label{Fig.7}
\end{figure*}

The field and temperature dependencies of the obtained thermodynamic parameters for different protocols are shown in Figs.(\ref{Fig.5}, \ref{Fig.6}, \ref{Fig.7}) for some values of $c$, $t_0$ and $\beta$. All magnetizations are continuous functions of field and temperature but susceptibilities can have jumps. The breaks, jumps and merging points in Figs. (\ref{Fig.5}, \ref{Fig.6}, \ref{Fig.7}) are associated with the loop's characteristic fields and the temperatures at which line $h=const.$ crosses the graphs of these fields in Fig. \ref{Fig.2}, i. e. the solutions to equations
\[
h_k(t_k)=h, \qquad k=c,d,e,f.
\]
Also we should add to them $t_{f0}$ of Eq.(\ref{eq:9}), $t_s$ and $3t_s$. For $c>c_0$, $h<h_{fmax}$ the relevant $t_f$ is the largest root of the above equation.
There is one more characteristic field $h_{3s}$ for $TR$ process defined as
\[
3t_s(h_{3s})=t. 
\]
Above it $m_{TR}=m_u(h=0)$ and does not depend on $h$, see Fig.\ref{Fig.5}a and Eqs. (\ref{eq:26}, \ref{eq:16}).

The jumps of $\chi_{ZFC}$ at $h_e$ and $t_e$ in Figs. (\ref{Fig.5}b, \ref{Fig.7}) result from the inhomogeneous approach of $m(n,t,h)$ to $m_0$ when $n \to 1$ and those of $\chi_{TR}$, $\chi_{IR}$ at $t=0$ in Fig. \ref{Fig.7} are the consequence of $n_d \to 2/3$ at $t \to 0$. So the influence of external field makes system to be trapped in  $n=2/3$ state at $t \to -0$ while at $t \to +0$ the $n=1$ homogeneous state is realized. Hence $\chi_{TR}=\chi_{IR}=\chi_u(h=0)=1/at_0$ at $t \to -0$, see Eqs. (\ref{eq:17}, \ref{eq:18}, \ref{eq:28}, \ref{eq:29}), and at $t \to +0$ $\chi=\chi_0=1/2at_0$, cf. Eq. (\ref{eq:14}).

Also $m_{TR}=m_{IR}=m_u(h=0)$ vanish at $t \to -0$. From Eqs. (\ref{eq:6},  \ref{eq:15}, \ref{eq:26}, \ref{eq:27}) we get at small $t$
\[
m_{IR,TR}  \approx \left[ {\frac{{ - tt_0^2 }}{{8\left( {3 + 2\beta } \right)}}} \right]^{3/2}
\]

Here we should note that the experimental observation of the obtained quasi-static magnetic susceptibilities can be achieved through the application of a small ac field with sufficiently small frequency. On the temperature and field intervals where the system evolves through the succession of states on the upper or lower boundary of the loop it is the only way to get the definite susceptibility value as the ordinary quasi-static susceptibility as field derivative of $m(h)$ curve does not exist in these regions. Indeed, the increasing of field in the state on the upper boundary brings the system inside the loop along the $m=m(n, t, h)$ curve but decreasing of it makes the system follow the loop outline. So the left and the right derivatives of the quasi-static $m(h)$ curve are different. Meanwhile the slow and small ac field will oscillate along the $m=m(n, t, h)$ curve only and we get the susceptibility defined in the obtained above expressions.
		 
Yet near $T_g$ the extremely slow relaxation can prevent the achievement of quasi-static regime at the laboratory frequencies. This may explain the absence of experimental data for susceptibilities similar to that of Figs.(\ref{Fig.5}b, \ref{Fig.7}). However the present results on the magnetizations in different protocols conform qualitatively to the experiments.  The field dependencies of magnetizations (polarizations) in Fig. \ref{Fig.5}a agree well with the experimental data of Refs. \cite{13}-\cite{16}. 

The rather unexpected result of the present theory is the prediction that  
\[
\mathop {\lim }\limits_{h \to 0} m_{TR,IR} \left( h \right) \ne 0
\] 
at $t<0$, see Fig. \ref{Fig.5}a. Indeed, from Eqs. (\ref{eq:20}, \ref{eq:21}, \ref{eq:22}, \ref{eq:26}) it follows that for $h \to 0$ the system is trapped in $n_s = 2/3$ state which is magnetized. Yet the only definite experimental witness in favor of this result we have found in the $m_{TR}$ data of Ref \cite{15}. Also $\chi_{FC}=\chi_{TR}$ at $h=0$ coincide with $\chi$ of $n=2/3$ state, cf. Eqs.(\ref{eq:24}, \ref{eq:28}). According to Eqs.(\ref{eq:11}, \ref{eq:13}, \ref{eq:17}) $\chi_{FC}=\chi_{TR}$ at $h=0$ merge at the value $\chi(1,0.5,t)$, see Eq.(\ref{eq:18}).

The temperature dependencies of $m_{FC}$ in Fig. \ref{Fig.6} exhibit two essentially different type of behavior, in Fig. \ref{Fig.6}a $m_{FC}$ grows monotonously at $t$ decreasing while in Fig. \ref{Fig.6}b it has a downward kink at $t=t_f$. This happens due to the different $c$ values, $c>\beta/(2+\beta)$ in the first case and $c<\beta/(2+\beta)$ in the second one. The both types of $m_{FC}$ behavior are observed in real spin glasses and ferroelectric relaxors, that of  Fig. \ref{Fig.6}a was seen in Refs. \cite{14}, \cite{16}-\cite{20} and  that of  Fig. \ref{Fig.6}b was registered in Refs. \cite{2}, \cite{15}, \cite{21}, \cite{24}. In both cases there is a reasonable qualitative agreement with the present theoretical results. 

The behavior of $m_{FC}$ and $m_{ZFC}$ for pure glass phase in Fig. \ref{Fig.6} can be compared with that for the case when glass phase appears as intermediate one between para- and ferro-phase (see Fig. 12 in Ref. \cite{8}). The qualitative difference is the jump of $m_{ZFC}$ in stronger fields in the last case. It can not appear in pure glass phase so the presence of this jump indicates that the glass phase is followed by the ferro-phase at lower temperatures.

\section{Temperature evolution after standard protocols}
\label{sec:4}
\begin{figure}[htp]
\centering
\includegraphics[height=7cm]{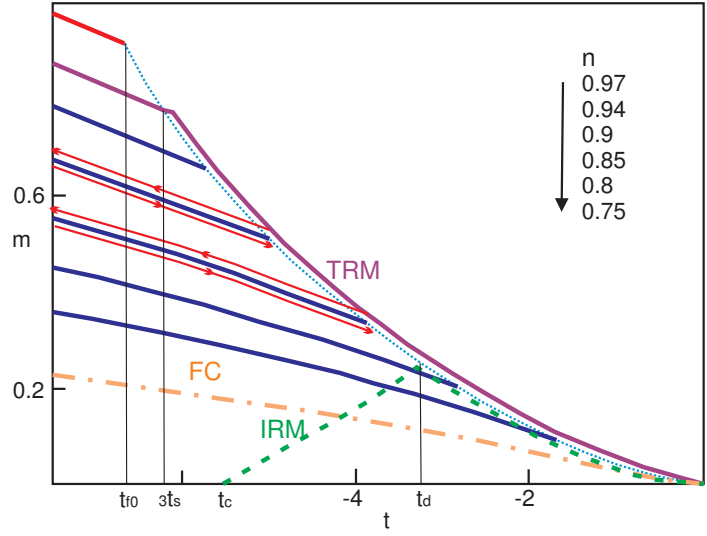}
\caption{(color online) Temperature dependencies of $M_{FC}(h=0)$, $M_{TR}(h=0.3a)$, $M_{IR}(h=0.9a)$ and zero-field magnetizations of metastable states for $\beta =5$, $c=0.9$, $t_0=1$. Directed lines shows the evolution of $TR$ state under cooling-heating cycles in zero field.}\label{Fig.8}
\end{figure}

Now we can consider the further evolution of $IR$ and $TR$ states under heating or cooling in zero field. To do this we turn to the Fig. \ref{Fig.8} where the temperature dependencies of $m_{TR}$ and $m_{IR}$ for $h=0.3a$ are shown along with those of several metastable states at $h=0$. Here $m_{TR}$ is the magnetization of the $n_s=0.98$ state below $t=3t_s(h=0.3a)=-6.2$ and above $3t_s$ it acquires the zero-field $m$ value on the upper branch of the loop (the $x=2$ stability boundary). Apparently the quasi-static heating will make $m_{TR}$ to follow its curve but under the quasi-static cooling the $TR$ state above $3t_s$ will join some metastable curve as shown in Fig. \ref{Fig.8} by the directed lines. If after that we heat the system to a higher $t$ and repeat the cooling-heating cycle we find that system evolve in it along the metastable state with a lower $m(n,h,t)$. Thus repeating the "cooling - heating for higher $t$" cycles one can get a "staircase" of magnetization's curves, explicitly demonstrating the existence of numerous metastable states in spin-glass phase.

Such experiments with zero-field temperature cycling of $TR$ state were made for iron-nitride fine particles and canonical spin-glass, CuMn alloy \cite{2} and for PLZT ceramic relaxor \cite{22} with the results qualitatively similar to that of Fig. \ref{Fig.8}. We may also note that Fig. \ref{Fig.8} depicts actually the $TRM$'s for several different fields as for another fields they will be composed just of another  matestable $n_s$ branches joining the $x=2$ boundary at another $3t_s$. The $TR$ polarizations for different fields was measured in ferroelectric relaxor $Cd_2Nb_2O_7$ \cite{16} and the results are quite similar to that of Fig. \ref{Fig.8}.

The $m_{IR}(h=0.9a)$ curve in Fig. \ref{Fig.8} crosses at $t_c(h=0.9a)=-5.6<t<t_d(h=0.9a)=-3.2$ the $m(n,h,t)$ lines with $0.5<n<n_d(t_d(h=0.9a))=0.82$. So in this temperature region the quasi-static zero-field heating or cooling of $IR$ state makes system join one of these metastable states (and not follow $m_{IR}(h=0.9a)$ curve). At $t>t_d(h=0.9a)=-3.2$ the $IR$ state is on the upper branch of the loop and the situation is similar to that of $TR$ state considered above. 
 
\begin{figure*}
\centering
\includegraphics{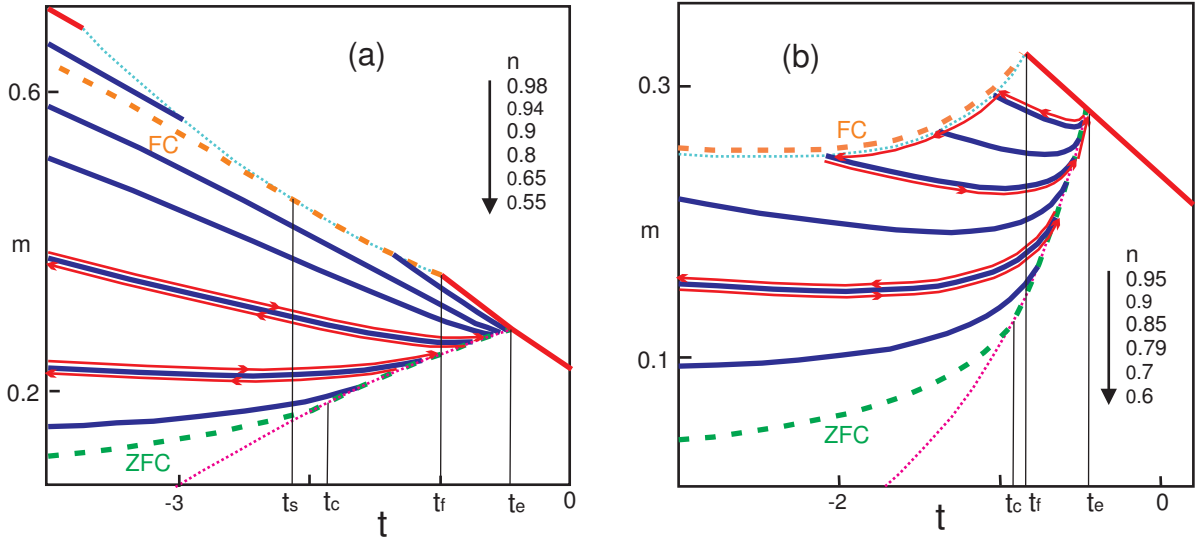}
\caption{(color online)(a) Temperature dependencies of $M_{FC}(h=0.3a)$,  $M_{ZFC}(h=0.3a)$ and the magnetizations of several metastable states at $h=0.3a$ for $\beta =5$, $c=0.9$, $t_0=1$  ($c>\beta/(2+\beta)$). Directed lines shows the evolution of $ZFC$ state under cooling-heating cycles in the field $h=0.3a$. (b) Temperature dependencies of $M_{FC}(h=0.1a)$, $M_{ZFC}(h=0.1a)$ and the magnetizations of several metastable states at $h=0.1a$ for $\beta =5$, $c=0.5$, $t_0=0.1$ ($c<\beta/(2+\beta)$). Directed lines show the evolution of $ZFC$ and $FC$ states under cooling-heating cycles in the field $h=0.1a$.}\label{Fig.9}
\end{figure*}

Similarly we can consider the further cooling or heating evolution of $ZFC$ and $FC$ states in the constant field achieved in these protocols. Fig. \ref{Fig.9}a shows the temperature dependencies of $m_{ZFC}$ and $m_{FC}$ for $h=0.3a$  along with those of several metastable states at the same field. Here $m_{ZFC}$ at $t<t_c(h=0.3a)=-1.9$ is the magnetization of the $n=0.5$ state in $h=0.3a$ and at $t_c(h=0.3a)<t<t_e(h=0.3a)=-0.46$ it is that of the lower loop's branch in the same field. Again simple heating makes the system follow the $m_{ZFC}$ curve, while the cooling-heating cycles with the subsequently rising upper turning points result in staircase of magnetization curves as directed lines in Fig.\ref{Fig.9}a show.

The $m_{FC}(h=0.3a)$ at $t<t_s(h=0.3a)=-2.1$ is the magnetization of the $n_s=0.97$ state in $h=0.3a$ and at $t_s(h=0.3a)<t<t_f(h=0.3a)=-1$ it is that of the upper loop's branch in the same field. The cooling in the field results also in the following the $m_{FC}$ curve but on heating above $t_s$ the system joins one of the metastable states with $n_s<n<1$. The last effect could hardly be observed for the case $c>\beta/(2+\beta)$ in Fig. \ref{Fig.9}a. It is much more pronounced for $c<\beta/(2+\beta)$ as in Fig. \ref{Fig.9}b. Here the temperature hysteresis cycles can be seen between $m_{ZFC}$ and $m_{FC}$.

We should note that the staircases of metastable magnetization curves starting from $TR$ and $ZFC$ states are also obtained in the random magnet having the intermediate ferromagnetic phase between paramagnetic and glassy ones \cite{1}.
In this case the boundary of $ZFC$ staircase differs by a steeper rise from that of Fig. \ref{Fig.9}a in conformity with the results of Ref. \cite{8}.
 
\section{Conclusions}
\label{sec:5}
The present results allow for the following conclusions:

1. The reasonable agreement of the present theory with the experiments implies the validity of the suggested in Refs. \cite{8}, \cite{9} mechanism of the phase-space separation in glassy phases. Yet further studies are needed to decide if the true nonergodicity sets in them so the description in terms of multiple "metastable states" can be extended beyond the present laboratory time scales. 

2. The present theory demonstrates the possible form which the results of future microscopic theory could have to describe the irreversible phenomena in random media on (at least) moderate yet macroscopic time scales.

3. The basic feature of the present approach is the filling of the interior of hysteresis loop with the magnetization curves of metastable states. The presence of strongly irreversible behavior only inside such loops seems to be the universal property of real dipole- and spin-glasses which should be necessary present in future theoretical developments.

4. One more essential result is that the states at the loop's sides are on the verge of their stability. So the random magnets and ferroelectrics are in the regions of self-organized criticality when they traverse these stability boundaries under some field and temperature variations. Underlying physical picture suggests here permanent upturns (avalanches) of fractal sets of local spins (dipoles) which can be probably registered in experiments. 

5. It is hard to say if the nonvanishing $m_{FC}$ and $m_{TR}$ at zero field and the susceptibility jumps are just the properties of the present model or the universal features of glassy phases. It seems worthwhile to study these issues in real and numerical experiments.

	I gratefully acknowledge useful discussions with V.B. Shirokov, M.P. Ivliev,  E.D. Gutlianskii and V.P. Sakhnenko.


%
%
\end{document}